# Perturbation of Nuclear Decay Rates During the Solar Flare of 13 December 2006

Jere H. Jenkins[1] and Ephraim Fischbach[1]

[1]*Physics Department, Purdue University, West Lafayette, IN, USA*

Correspondence to: Ephraim Fischbach[1] Correspondence and requests for materials should be addressed to E.F. (Email: Ephraim@physics.purdue.edu).

**Recently, Jenkins, et al. have reported the detection of correlations between fluctuations in nuclear decay rates and Earth-Sun distance, which suggest that nuclear decay rates can be affected by solar activity. In this paper, we report the detection of a significant decrease in the decay of $^{54}$Mn during the solar flare of 13 December 2006, whose x-rays were first recorded at 02:37 UT (21:37 EST on 12 December). Our detector was a 1 μCi sample of $^{54}$Mn, whose decay rate exhibited a dip coincident in time with spikes in both the x-ray and proton fluxes recorded by the GOES-10 and 11 satellites. A secondary peak in the x-ray and proton fluxes on 17 December at 12:40 EST was also accompanied by a coincident dip in the $^{54}$Mn decay rate. These observations support the claim by Jenkins, et al. that nuclear decay rates vary with Earth-Sun distance.**

Solar flares are periods of increased solar activity, and are often associated with geomagnetic storms, solar radiation storms, radio blackouts, and similar effects that are experienced here on Earth. It has been speculated that the increased activity associated with solar flares may also produce a short-term change in the neutrino flux detected on Earth.[1,2,3,4,5,6] To date, there appears to be no compelling experimental evidence of an association between neutrino flux and solar flares,[1,2,4,6] and this is due in part to the relatively low neutrino counting rates available from even the largest conventional detectors.

The object of the present paper is to use data we obtained during the solar flare of 13 December 2006 to suggest that neutrinos from the flare were detected via the change they induced in the decay rate of $^{54}$Mn. The present paper supports the work of Jenkins, et al. who present evidence for a correlation between nuclear decay rates and Earth-Sun distance[7]. Taken together, these papers suggest that nuclei may respond to changes in solar activity, possibly arising from changes in the flux of solar neutrinos reaching the Earth.

The apparatus that was in operation during the solar flare is described in detail in the Supplemental Material. During the course of the data collection in the Physics building at Purdue University which extended from 2 December 2006 to 2 January 2007, a solar flare was detected on 13 December 2006 at 02:37 UT (21:37 EST on 12 December) by the Geostationary Operational Environmental Satellites (GOES-10 and GOES-11). Spikes in the x-ray and proton fluxes were recorded on all of the GOES satellites.[8] The x-ray data from this X-3 class solar flare are shown in Figures 1-3 along with the $^{54}$Mn counting rates: In each 4 hour live-time period (~4.25 hours real-time) we recorded

~2.5x10$^{7}$ 834.8 keV γ-rays with a fractional $1/\sqrt{N}$ statistical uncertainty of ~2x10$^{-4}$. Each data point in Figs. 1-3 then represents the number of counts in the subsequent 4 hour period, which are normalized in Figs. 2 and 3 by the number of counts $N(t)$ expected from a monotonic exponential decay, $N(t) = N_0 \exp(-\lambda t)$, with λ=0.002347(2)d$^{-1}$ determined from our December data. We see from Figs. 1-3 that, to within the time resolution offered by the 4 hour width of our bins, the $^{54}$Mn counting rates exhibit a dip which is coincident in time with the spike in the x-ray flux which signalled the onset of the solar flare. Although a second x-ray peak on 14 December at 17:15 EST corresponds to a relatively small dip in the $^{54}$Mn count rate, a third peak on 17 December at 12:40 EST is again accompanied by an obvious dip in the $^{54}$Mn counting rate, as seen in Figs.1-3. The fact that some x-ray spikes in these and other data sets are not accompanied by correspondingly prominent dips in the $^{54}$Mn data may provide clues to the underlying mechanisms that produce these solar events. Conversely, peaks or dips in the $^{54}$Mn data not accompanied by visible x-ray spikes may correspond to other solar events, or events on the opposite side of the Sun, which are possibly being detected via neutrinos. In particular, the dip on 22 December (09:04 EST) was coincident in time with a severe solar storm,[9] but did not have an associated x-ray spike. Additionally, when more data become available, we may find that neutrino oscillations and other time-dependent phenomena suggested by the data of Ref. 7 may play a significant role in solar flares as well.

Before considering more detailed arguments in support of our inference that the $^{54}$Mn count rate dips are due to solar neutrinos, we address the question of whether the coincident fluctuations in the decay data and the solar flare data could simply arise from statistical fluctuations in each data set. Referring to Fig.3, we define the dip region in the decay data as the 84 hour period (encompassing our runs 51-71 inclusive) extending between 11 December 2006 (17:52 EST) and 15 December 2006 (06:59 EST). The measured number of decays $N_m$ in this region can then be compared to the number of events $N_e$ expected in the absence of the observed fluctuations, assuming a monotonic exponential decrease in the counting rate. Since the systematic errors in $N_e$ and $N_m$ are small compared to the statistical uncertainties in each, only the latter are retained and we find,

$$N_e - N_m = (7.51 \pm 1.07) \times 10^5, \qquad (1)$$

where the dominant contributions to the overall uncertainty arise from the $\sqrt{N}$ fluctuations in the counting rates. If we interpret Eq. 1 in the conventional manner as a ~7σ effect, then the formal probability of such a statistical fluctuation in this 84 hour period is ~3x10$^{-12}$. Evidently, including additional small systematic corrections would not alter the conclusion that the observed fluctuation in runs 51-71 is not likely a purely statistical effect.

We next estimate the probability that a solar flare would have occurred during the same 84 hour period shown in Fig. 3. The frequency of solar radiation storms varies with their intensities, which are rated on a scale from S1(Minor) to S5(Extreme).[8] The 13 December 2006 event was rated as S2 (Moderate), and S2 storms occur with an average frequency of 25 per 11 year solar cycle.[8] In total, the frequency of storms with intensity $\geq$S2 is ~39 per 11 year solar cycle, or $9.7\times10^{-3}$, and hence the probability of a storm occurring at any time during the 84 hour window in Fig. 3 is ~$3.4\times10^{-2}$. Evidently, if the x-ray and decay peaks were uncorrelated, the probability that they would happen to coincide as they do over the short time interval of the solar flare would be smaller still, and hence a conservative upper bound on such a statistical coincidence occurring in any 84 hour period is ~$(3\times10^{-12})(3\times10^{-2}) \approx 1\times10^{-13}$. Since a similar analysis would apply to the coincident peak and dip at 12:40 EST on 17 December, the probability that random fluctuations would produce two sets of coincidences several days apart is negligibly small, and hence we turn to consider other possible explanations for the data in Figs. 1-3.

As stated before, solar flares are known to produce a variety of electromagnetic effects on Earth, including changes in the Earth's magnetic field, and power surges in the electric grids. It is thus conceivable that the observed dips in the $^{54}$Mn counting rate could have arisen from the response of our detection system (rather than the $^{54}$Mn atoms themselves) to the solar flare. In the Supplemental Material we demonstrate that the observed dip in the $^{54}$Mn counting rate coincident with the solar flare at 21:37 EST on 12 December 2006 are not likely the result of a conventional electromagnetic or other systematic effect. We therefore turn to the possibility that this dip was a response to a change in the flux of solar neutrinos during the flare, as implied by the analysis of Jenkins, et al.[7].

We begin by noting that the x-ray spike occurred at ~21:40 EST, approximately 4 hours after local sunset, which was at ~17:21 EST on 12 December 2006. As can be seen from Fig. 4, the neutrinos (or whatever agent produced this dip) had to travel ~9,270 km through the Earth before reaching the $^{54}$Mn source, and yet produced a dip in the counting rate coincident in time with the peak of the x-ray burst. Significantly, the monotonic decline of the counting rate in the 40 hours preceding the dip occurred while the Earth went through 1.7 revolutions, and yet there are no obvious diurnal or other periodic effects. These observations support our inference that this effect may have arisen from neutrinos, or some neutrino-like particles, and not from any conventionally known electromagnetic effect or other source, such as known charged particles.

If the detected change in the $^{54}$Mn decay rate was in fact due to neutrinos then one implication of the present work is that radioactive nuclides could serve as real-time neutrino detectors for some purposes. In principle, such "radionuclide neutrino detectors" (RNDs) could be combined with existing detectors, such as Super-Kamiokande, to significantly expand our understanding of both neutrino physics and solar dynamics. One potentially interesting application of such an RND would be to the detection of the relic neutrino sea remaining from the big bang. It is estimated[11] that the present relic neutrino density is ~56 $cm^{-3}$/species, which is ~336 $cm^{-3}$ in total. If we assume that the velocity of the Earth relative to the relic neutrino sea is ~370 $km\cdot s^{-1}$, then the resulting flux of neutrinos incident on an RND would be ~$1\times10^{10} cm^{-2}s^{-1}$. This is comparable to the estimated solar flux, ~$6\times10^{10} cm^{-2}s^{-1}$ and potentially comparable to the fluctuation in the neutrino flux detected during the solar flare period. This use of RNDs may suggest a realistic experiment capable of detecting the relic neutrino sea.

**Acknowledgements** The authors are indebted to A.M. Hall and M.A. Fischbach for their many contributions during the early stages of this collaboration, and to J. Buncher, D. Krause, and J. Mattes for critically reading this manuscript. We also express our appreciation to D. Anderson, B. Craig, J. Gruenwald, J. Heim, A. Longman, E. Merritt, T. Mohsinally, D. Mundy, J. Newport, B. Revis, J. Schweitzer, M. Sloothaak, and A. Treacher for their assistance and for many helpful conversations. The work of E.F. was supported in part by the U.S. Department of Energy under Contract No.DE-AC02-76ER071428.

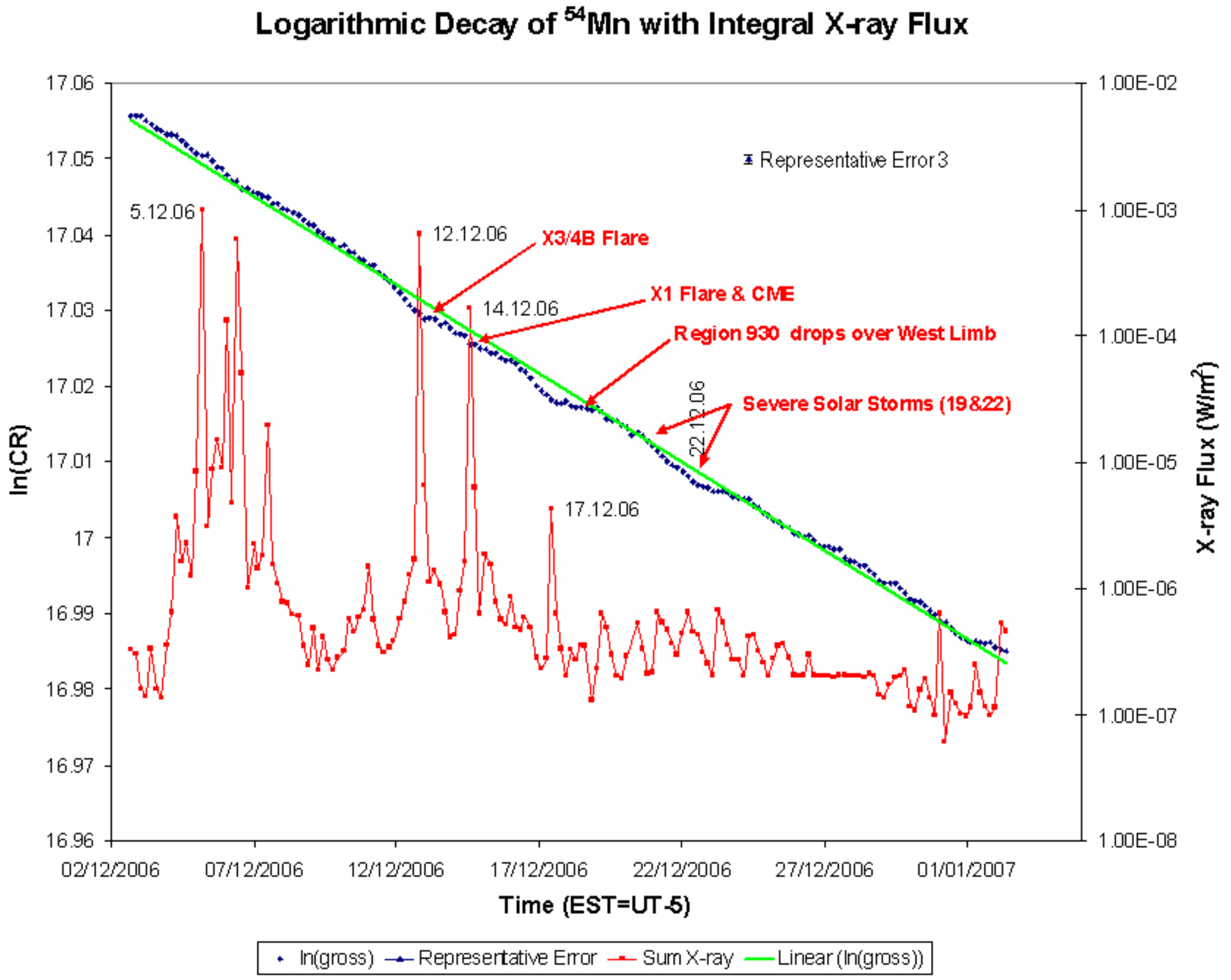

Figure 1. December 2006 $^{54}$Mn data, and GOES-11 x-ray data, both plotted on a logarithmic scale. For $^{54}$Mn, each point represents the natural logarithm of the number of counts ~$2.5\times10^{7}$ in the subsequent 4 hour period, and has a $\sqrt{N}$ statistical error shown by the indicated error bar. For the GOES-11 x-ray data, each point is the solar x-ray flux in W/m$^2$ summed over the same real time intervals as the corresponding decay data. The solid line is a fit to the $^{54}$Mn data, and deviations from this line coincident with the x-ray spikes are clearly visible on 12/12 and 17/12. As noted in the text, the deviation on 22/12 was coincident with a severe solar storm, with no associated flare activity.[9] The dates for other solar events are also shown by arrows.

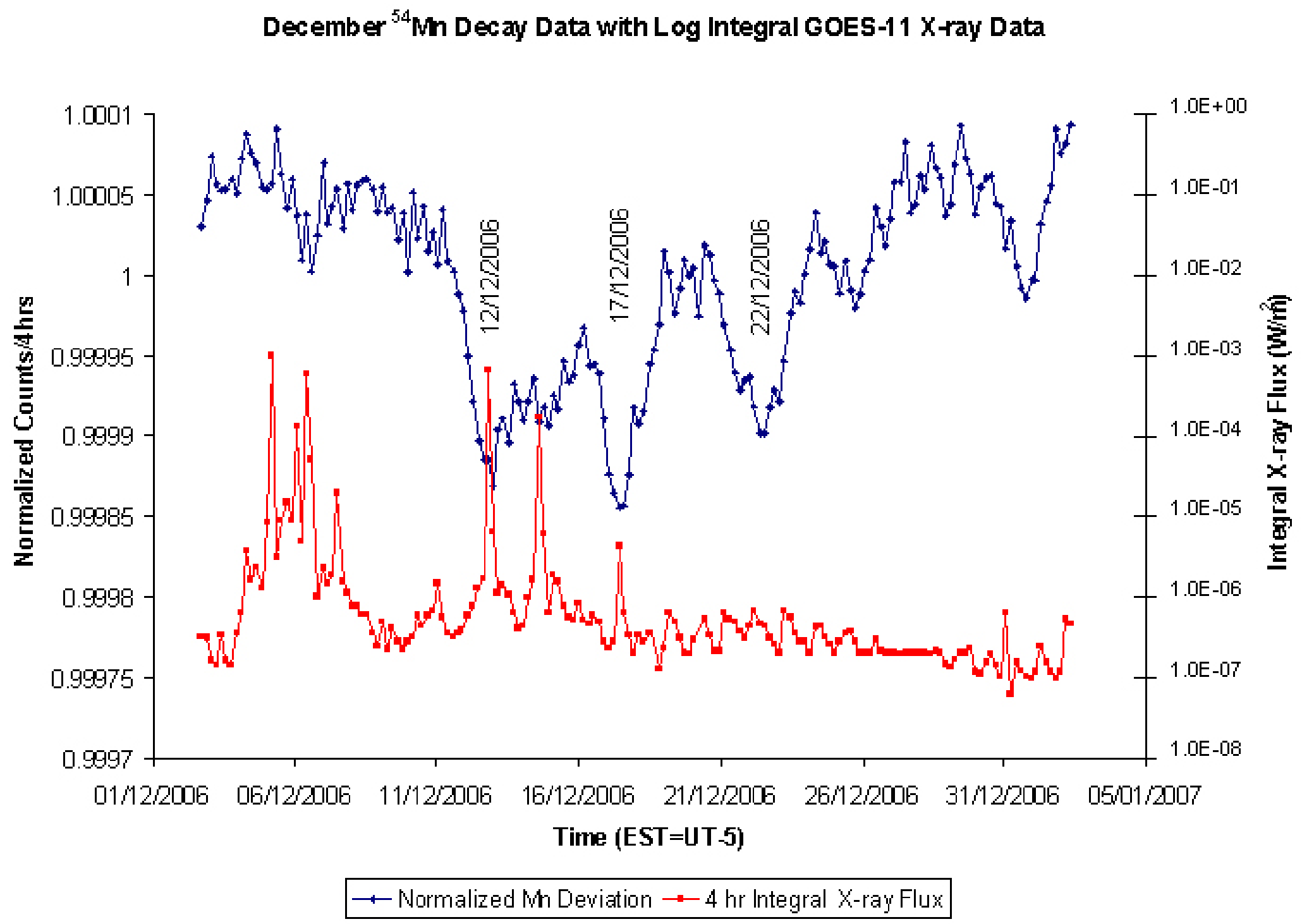

Figure 2. Normalized December 2006 $^{54}$Mn decay data along with GOES-11 x-ray data on a logarithmic scale. For $^{54}$Mn, each point represents the number of counts in the subsequent four hour period normalized to the average decay rate (see text), and has a fractional $\sqrt{N}$ statistical uncertainty of ~2x10$^{4}$. For the GOES-11 x-ray data, each point is the solar flux in W/m$^{2}$ summed over the same real-time intervals. The 12 December peak in the x-ray flux occurred at ~21:37 EST.

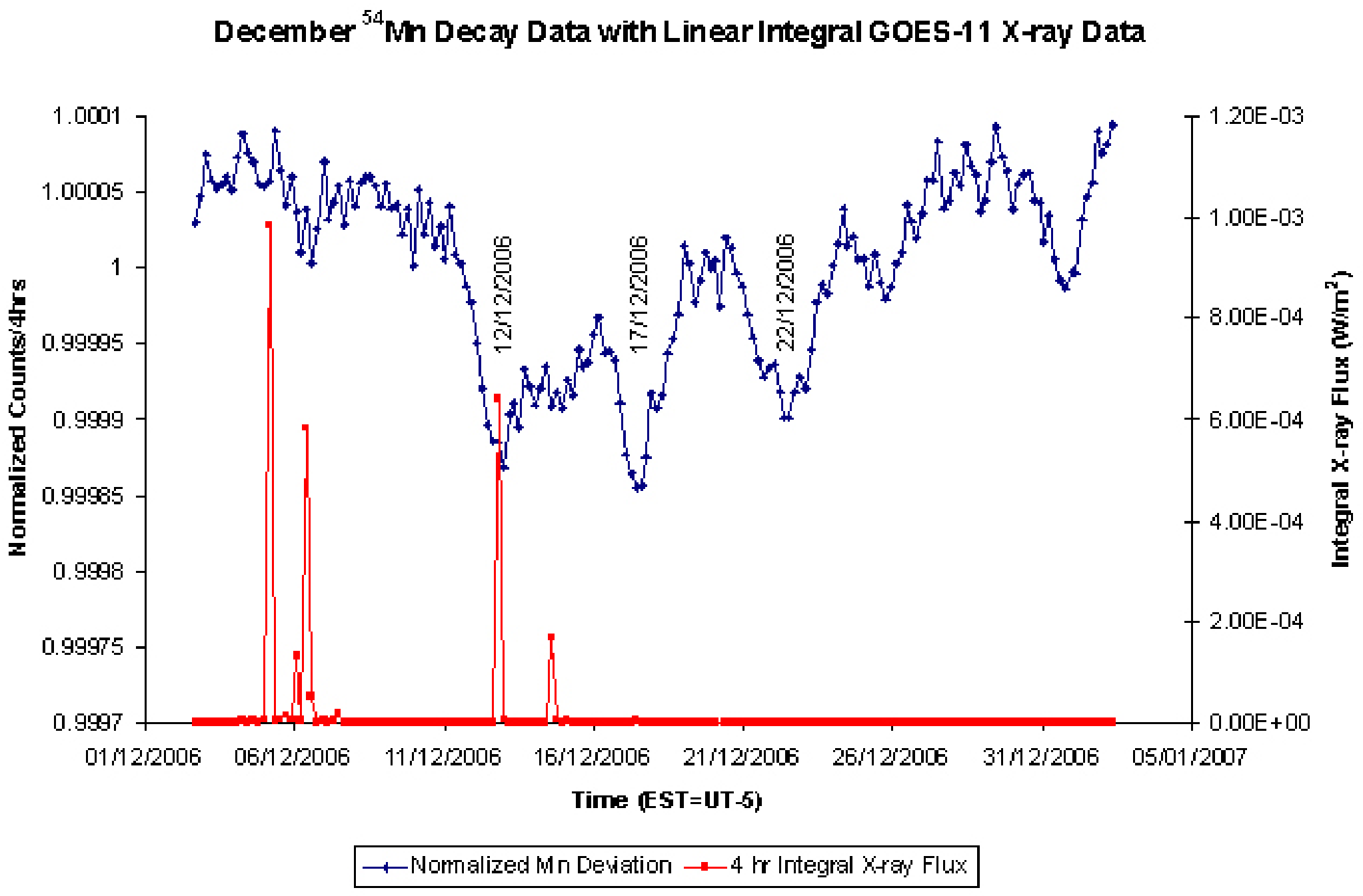

Figure 3. Normalized December 2006 $^{54}$Mn decay data along with GOES-11 x-ray data on a linear scale. The solar flare at ~21:37 EST on 12 December is clearly visible, along with the precursor count-rate decline that precedes it. See text and captions to Figs. 1 and 2.

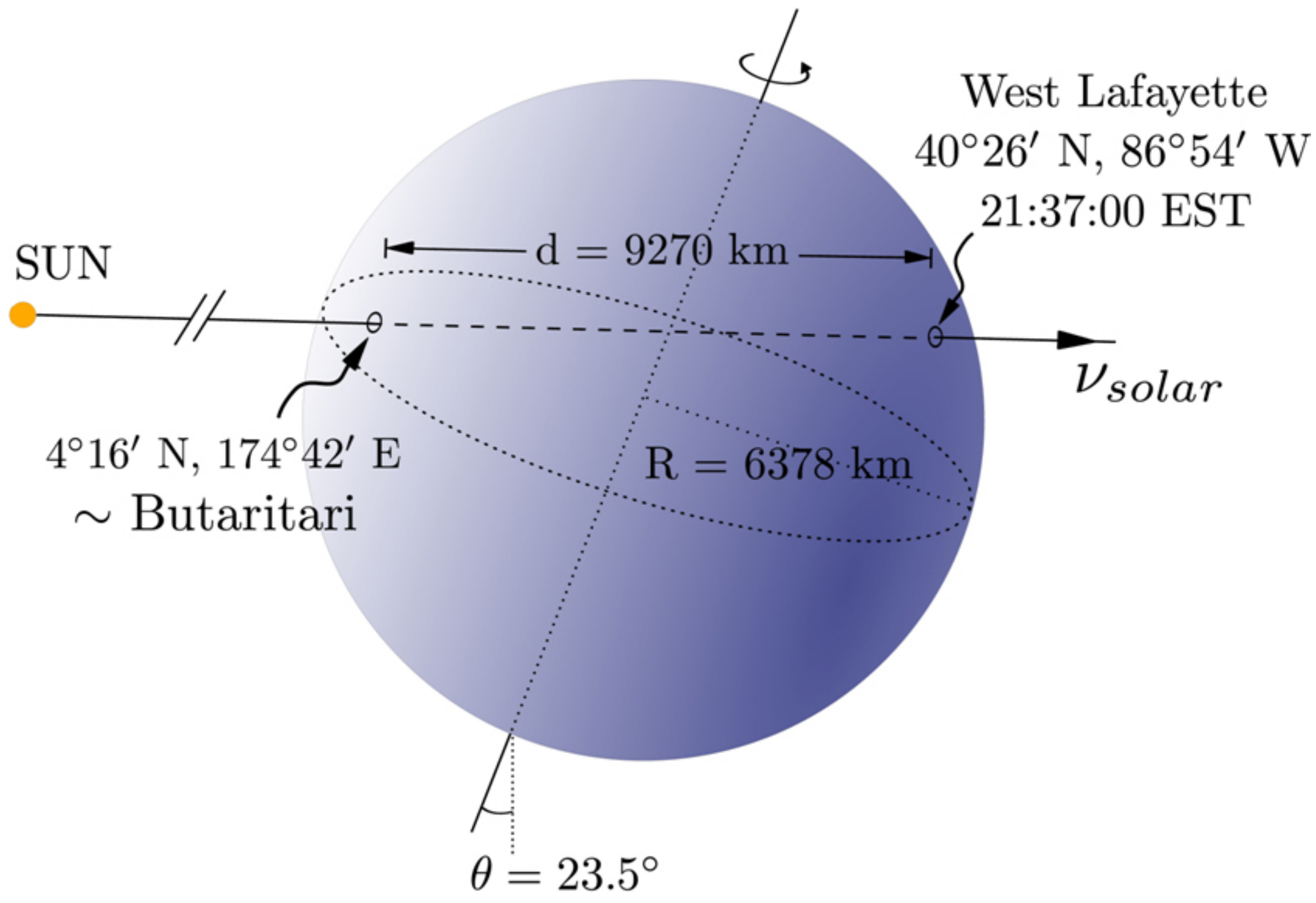

Figure 4. Trajectory of neutrinos from the solar flare of 12 December 2006. The neutrinos would have entered the Earth near Butaritari, in the Pacific Ocean, and travelled ~9270 km through the Earth before the coincident minimum in the count-rate was detected in West Lafayette, Indiana.

# Supplementary Discussion

### The experimental setup

The apparatus that was in operation during the solar flare was a ~1 μCi sample of $^{54}$Mn attached to the front of a Bicron 2x2 inch NaI(Tl) crystal detector (model number 2M2/2-X), which was connected to an Ortec photomultiplier (PMT) base with pre-amplifier. An Ortec 276 spectroscopy amplifier was used to analyze the pre-amplifier signal, and this was connected to an Ortec Trump(R) PCI card running Ortec's Maestro32® MCA software. The system recorded the 834.8 keV γ-ray emitted from the de-excitation of $^{54}$Cr produced from the K-capture process $^{54}\mathrm{Mn} + e^{-} \rightarrow {}^{54}\mathrm{Cr} + \nu_e$. The detector and $^{54}$Mn sample were shielded on all sides by lead bricks, except at the end of the PMT base where a space was left to accommodate cables. The apparatus was located in a windowless, air-conditioned interior 1$^{st}$ floor room in the Physics building at Purdue in which the temperature was maintained at a constant 19.5(5)°C.

### Systematic sensitivity to environmental conditions

As stated in the text, solar flares are known to produce a variety of electromagnetic effects on Earth, including changes in the Earth's magnetic field, and power surges in the electric grids. It is thus conceivable that the observed dips in the $^{54}$Mn counting rate could have arisen from the response of our detection system (rather than the $^{54}$Mn atoms themselves) to the solar flare. The most compelling argument against this explanation of the $^{54}$Mn data is that the $^{54}$Mn decay rate began to decrease more than one day *before* any signal was detected in x-rays by the GOES satellites (see Figs. 1-3 in text). Since it is unlikely that any other electromagnetic signal would reach the Earth earlier than the x-rays, we can reasonably exclude any explanation of the $^{54}$Mn data in terms of a conventional electromagnetic effect arising from the solar flare. This is particularly true since the most significant impact on the geomagnetic field occurs with the arrival of the charged particle flux, several hours after the arrival of the x-rays.

We can further strengthen the preceding argument by examining in detail the response our detection system to fluctuations in line voltages. No unusual behavior was detected by either the Purdue power plant (private communication, Ron Porte, Purdue, 2007), or by the Midwest Independent Systems Operator (MISO) which also supplies power to Purdue (private communication, John Jenkins, MISO, 2007). MISO did in fact receive notification on 14 December 2006 of a “Geo-magnetic disturbance of K-7 magnitude” at 02:46 UT, but noted that there were no reported occurrences of excessive neutral currents during the time-frame of 10-18 December 2006. At Purdue, an alert would have been triggered had the line voltage strayed out of the range 115-126 V, and hence we can infer that the voltage remained within this range during the solar flare. Moreover, since the main effect of a power surge would have been to shift the $^{54}$Mn peak slightly out of the nominal region of interest (ROI) for the 834.8 keV γ-ray, this would have been noted and corrected for in the routine course of our data acquisition. No significant changes to either the peak shape or location were noted during this period.

We turn next to an examination of the effect of fluctuating magnetic fields on our detector system. Supplementary Figure 1 (SF1) exhibits the $A_p$ index for the Earth's magnetic field during December 2006,[1] along with the $^{54}$Mn counting rate. We see immediately that the sharp spike in the $A_p$ index at approximately 00:00 EST on 15 December 2006 occurred more than two days *after* the solar flare and the accompanying dip in the $^{54}$Mn counting rate, and hence was presumably not the cause of this dip. This conclusion can be further strengthened by the results of a series of measurements carried out in our laboratory, which are described in detail below. These results establish that our detection system was insensitive to applied magnetic fields that were more than 100 times stronger than the spike exhibited in Fig. SF1.

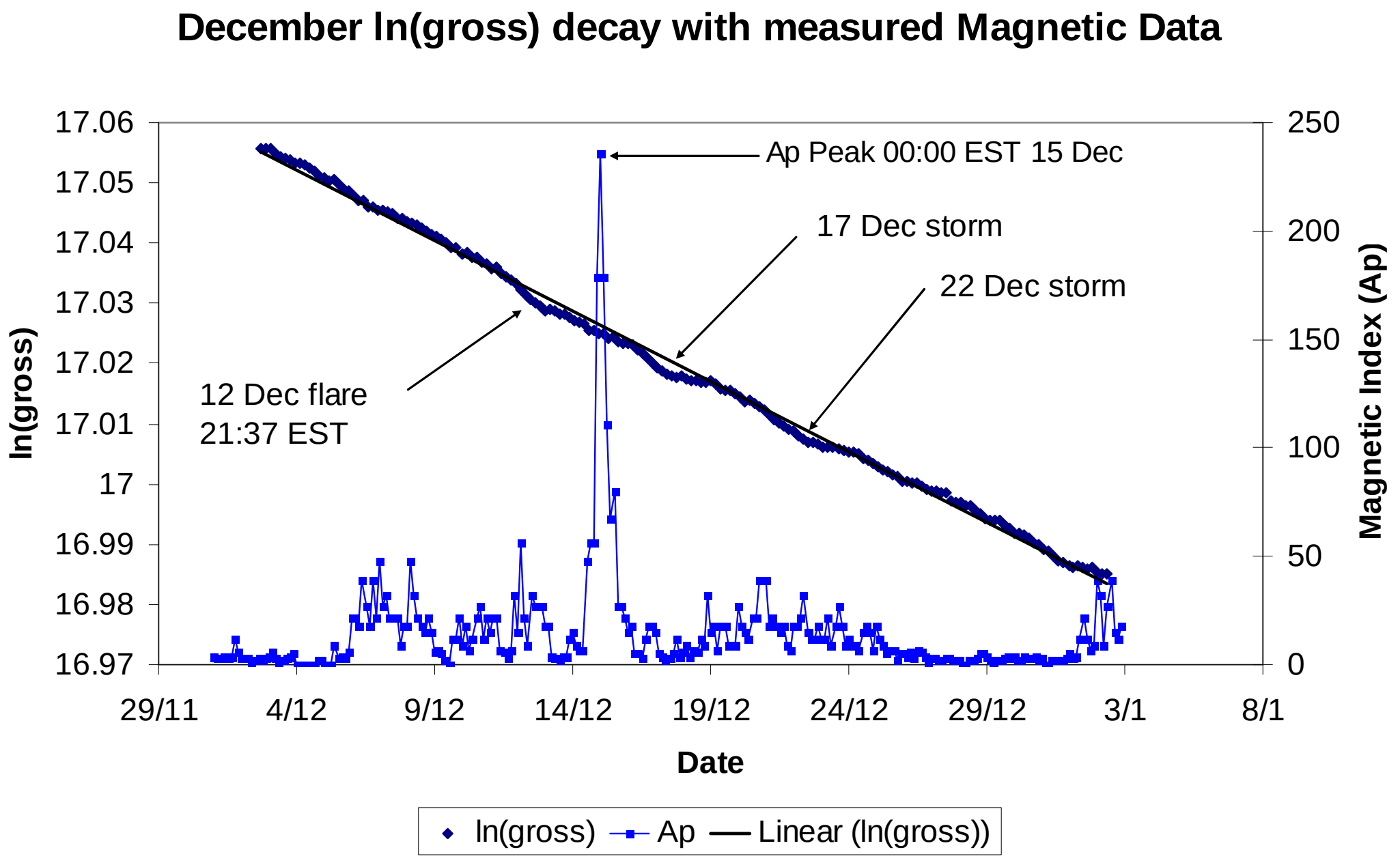

Supplementary Figure 1 | Fluctuations in the Earth's Magnetic Field in December 2006. The magnetic field fluctuations, which are characterized by the $A_p$ index, are plotted along with the natural logarithm of the $^{54}$Mn count rate. We note that the spike in the magnetic data on 15 December occurred ~2 days *after* the dip in the $^{54}$Mn count rate at 21:37 on 12 December.

To strengthen our conclusion that the observed dips in the $^{54}$Mn counting rate were not the result of the response of our detector to a change in the ambient magnetic field, we conducted a series of experiments which we outline here. The previously described Bicron NaI(Tl) crystal and the associated PMT were placed at the midplane of a pair of Helmholtz coils 33 cm in radius, obtained from a Welch Scientific Model 623A e/m apparatus, from which the e/m tube had been removed. Care was taken to orient the detector relative to the Earth's magnetic field exactly as it was during the counting period surrounding the solar flare. The calibration of the coils was such that a current of 0.1 Ampere produced a field of 0.196 Gauss at the midpoint of the two coils, and this calibration was independently verified by use of a Bell Model 610 Gauss meter. By orienting the coils to completely cancel the Earth's magnetic field, we determined from both the known current in the coils and the Gauss meter that the Earth's local magnetic field had a strength of 0.42(1) Gauss.

---

[1] NOAA/NWS Space Weather Pred. Ctr., http://www.sec.noaa.gov

To test the sensitivity of our apparatus to fluctuations in the magnetic field of the Earth, $|\vec{B}_\oplus|$, about its nominal value, we started with the data from Fig. SF1, which reported a spike with $A_p \approx 240$. This corresponds to a fluctuation in $|\vec{B}_\oplus| \leq 500\,\mathrm{nT}$ (1 nT =$10^{-5}$ Gauss), and hence we took $500\,\mathrm{nT} \cong 0.01|\vec{B}_\oplus|$ as a benchmark reference value. To avoid any drifts in the signal due to the decay of the sample over time, we replaced the $^{54}$Mn source with the longer lived $^{137}$Cs ($T_{1/2}$=30.07 years), whose decay photon at 661 keV is reasonably close to the 834 keV photon from $^{54}$Mn decay. Our results are shown in Fig. SF2, for an external field $|\vec{B}_\oplus|$ = 0, 0.42, and 0.85 Gauss. Note that the latter values for $|\vec{B}_\oplus|$ correspond respectively to ~100 and 200 times the maximum fluctuation measured during the flare by NOAA[2] (see Fig. SF1). We see from this figure that even fields this large produce no statistically significant fluctuations in the counting rates.

[2] NOAA/NWS Space Weather Pred. Ctr., http://www.sec.noaa.gov

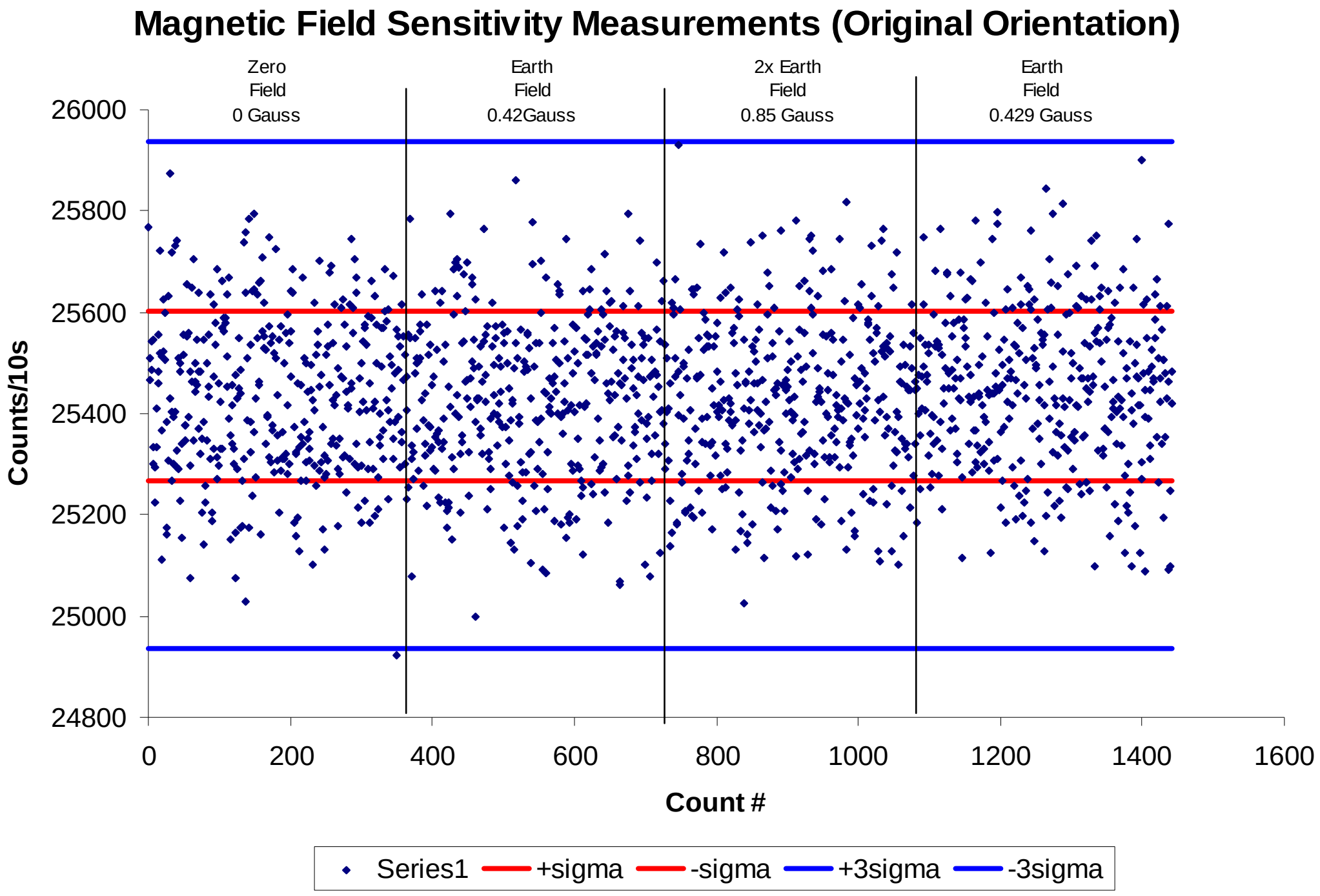

Supplementary Figure 2 | Dependence of Counting Rate on Applied Magnetic Field. Each point represents the number of $^{137}Cs$ counts in our detector during a 10 second run for the given magnetic field strength, which was varied between 0 and $2|\vec{B}_{\oplus}|$, where $|\vec{B}_{\oplus}|$=0.42 Gauss. As noted in the text, $2|\vec{B}_{\oplus}|$ represents a change approximately 200 times larger than any fluctuation measured during December 2006. For this set of data, the detector was oriented exactly as it was during the solar flare of 12 December 2006. The horizontal lines represent the $\pm 1\sigma$ and $\pm 3\sigma$ limits expected from the $\sqrt{N}$ fluctuations in the counting rates.

To avoid any possibility that the detector happened to be accidentally oriented in a direction which rendered it insensitive to the applied fields, the cylindrical detector was rotated by 45° about its symmetry axis, relative to its original orientation, and an additional set of runs was carried out. The results from these runs are presented in Fig. SF3, and again show no evidence for the dependence of the counting rate on the magnitude of the external field.

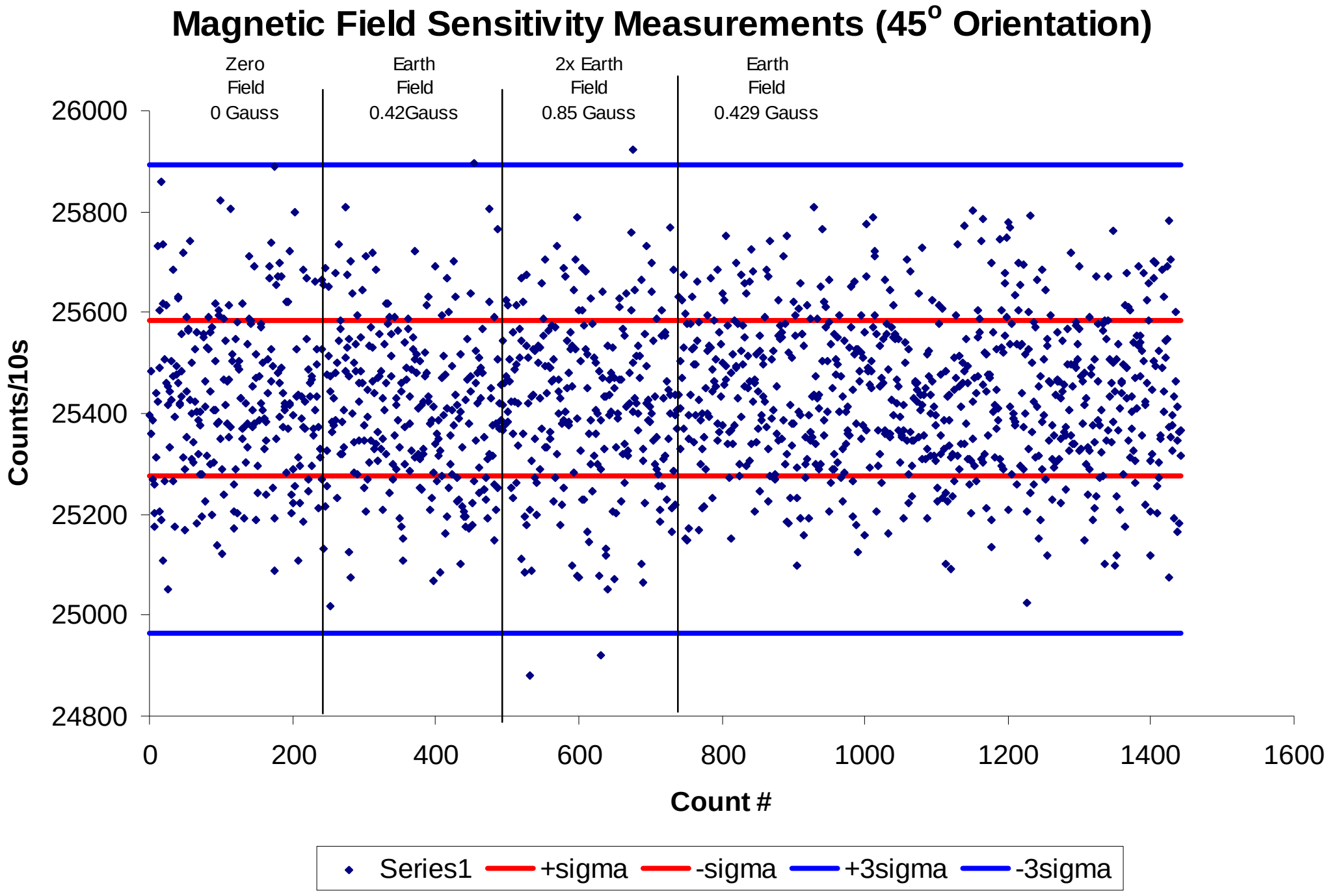

Supplementary Figure 3 | Data When the Detector was Rotated. In this set of runs, the cylindrical NaI(Tl) detector was rotated about its symmetry axis by 45°. See text and caption to Fig. SF2 for further details.

We conclude that since our apparatus is insensitive to fluctuations in the ambient magnetic field that are much larger than those observed following the flare of 12 December 2006, that even had the observed spikes coincided with the dip in the $^{54}$Mn data, they could not have accounted for the observed dip. In the end, this conclusion is not surprising, since the Bicron NaI(Tl) crystal and PMT are shielded against magnetic fields by 0.508 mm of mu-metal which makes up part of the detector housing.[3]

---

[3] Saint Gobain, 2000, Engineering Drawing 2M2/2-X, SA-04844 Rev. A